\documentclass[eps,prd,twocolumn,nofootinbib,showpacs]{revtex4}
\usepackage{graphicx}
\usepackage[colorlinks=true,linkcolor=blue,citecolor=blue,urlcolor=blue]{hyperref}

\begin{document}

\title{Analysis of $H \to J/\psi+\gamma$ up to Next-to-Next-to-Leading Order QCD Corrections}

\author{Wen-Yuan Li$^1$}

\author{Ting Sun$^1$}
\email[email:]{tingsuny@gzmu.edu.cn}

\author{Sheng-Quan Wang$^1$}
\email[email:]{sqwang@cqu.edu.cn}

\author{Jian-Ming Shen$^{2}$}

\author{Hua Zhou$^{3}$}

\author{Xing-Gang Wu$^4$}
\email[email:]{wuxg@cqu.edu.cn}

\author{Leonardo Di Giustino$^{5,6}$}

\address{$^1$Department of Physics, Guizhou Minzu University, Guiyang 550025, P.R. China}
\address{$^2$School of Physics and Electronics, Hunan University, Changsha 410082, P.R. China}
\address{$^3$School of Mathematics and Physics, Southwest University of Science and Technology, Mianyang 621010, P.R. China}
\address{$^4$Department of Physics, Chongqing University, Chongqing 401331, P.R. China}
\address{$^5$Department of Science and High Technology, University of Insubria, via Valleggio 11, I-22100, Como, Italy}
\address{$^6$INFN, Sezione di Milano-Bicocca, 20126 Milano, Italy}

\date{\today}

\begin{abstract}

The rare exclusive decay of the Higgs boson, $H \to J/\psi + \gamma$, is a crucial channel for assessing the Yukawa coupling of the charm quark. In this article, we examine this process up to the next-to-next-to-leading order (NNLO) in QCD utilizing the Principle of Maximum Conformality (PMC). The PMC offers a systematic approach to eliminate renormalization scale uncertainties by resumming non-conformal $\beta$ contributions into the QCD running coupling through the renormalization group equation (RGE). We determine a PMC scale of $Q_\star = 3.29\ \text{GeV}$, which reflects the low virtuality of the underlying QCD dynamics in the $H \to J/\psi + \gamma$ process. Notably, this is an order of magnitude smaller than the scale estimated using the conventional method, i.e., $\mu_r = m_H/2$. By removing the non-conformal $\{\beta_i\}$-terms from the perturbative QCD (pQCD) series, we observe that the PMC NLO QCD correction term is significantly enhanced, while the PMC NNLO QCD correction is suppressed. This indicates improved convergence of the pQCD series up to NNLO. Finally, we calculate the decay width $\Gamma(H \to J/\psi + \gamma) = 14.183^{+0.249}_{-0.347} \pm 0.022$ eV, where the first uncertainty arises from the factorization scale $\mu_\Lambda \in [1, 2]\ \text{GeV}$, and the second is due to the estimation of unknown higher-order terms using the Pad$\acute{e}$ approximant approach. The corresponding branching fraction is $\mathcal{B}(H \to J/\psi + \gamma) = 3.485_{-0.161}^{+0.152} \times 10^{-6}$.

\end{abstract}

\maketitle

\section{Introduction}
\label{sec:1}

In 2012, the ATLAS and CMS collaborations observed a new scalar boson with a mass of $125$ GeV~\cite{higgs1,higgs2,higgsmass1,sign2}. Given the unique properties of this new scalar particle and its perfect alignment with the predicted Higgs boson of the Standard Model (SM), it was soon identified as the Higgs boson~\cite{sign1,cms131,atlas131,atlascms131,atlascms13002}. The confirmation of the Higgs particle has firmly established the consistency and validity of the Standard Model as the fundamental theory capable of unifying the description of the strong and electroweak forces. The Yukawa coupling is a defining characteristic of the Higgs boson~\cite{Kagan:2014ila}. Currently, the most stringent constraint on the charm Yukawa coupling is $1.1<|\kappa_c|<5.5$, obtained by analyzing the $\bar{c}c$ jets in the final state by the CMS collaboration~\cite{CMS:2022psv}.

Measurements of the charm Yukawa coupling have been conducted via two different processes: the direct $H \rightarrow c\bar{c}$ decays and the exclusive decay process $H \to J/\psi+\gamma$. Although the branching ratio for the $H \to J/\psi+\gamma$ is small, it offers complementary information on the Yukawa coupling. Furthermore, the $H \to J/\psi+\gamma$ decay process provides a clean final state with the subsequent decay of $J/\psi$ into lepton pairs. Thus, the exclusive decay process $H \to J/\psi+\gamma$ is a promising channel for investigating the charm Yukawa coupling. This study is conducted by detecting the $\mu^+\mu^-\gamma$ final state with an integrated luminosity of $139$ fb$^{-1}$ at $\sqrt{s}=13$ TeV~\cite{ATLAS:2022rej}, yielding an upper limit on the branching ratio of $2.1\times 10^{-4}$ for the ATLAS collaboration.

Theoretical investigations of the $H \to J/\psi+\gamma$ decay process date back to 1980~\cite{Shifman:1980dk,Vysotsky:1980cz,Keung:1983ac}. The $H \to J/\psi+\gamma$ process is primarily studied using nonrelativistic QCD (NRQCD)~\cite{Bodwin:1994jh}. Significant efforts have been made to enhance the theoretical predictions for the $H \to J/\psi+\gamma$ process up to the order $\alpha_s v^0$ and $v^2$ of accuracy~\cite{Bodwin:2013gca,Bodwin:2014bpa,Konig:2015qat,Bodwin:2016edd,Zhou:2016sot,Bodwin:2017wdu}. Recently, the prediction for the $H \to J/\psi+\gamma$ process has been refined by including the $v^4$ relativistic correction~\cite{Brambilla:2019fmu}. QCD corrections at the next-next-leading-order (NNLO) have been calculated in Ref.~\cite{Jia:2024ini}. These significant improvements in QCD calculations enable predictions of Higgs properties with enhanced precision.

In the conventional procedure, the renormalization scale $\mu_r$ for the $H \to J/\psi+\gamma$ process is chosen based on the Higgs boson mass $m_H$ (e.g., $\mu_r=m_H/2=Q$), and theoretical uncertainties are estimated by varying the scale in a specified range (e.g., $\mu_r\in[Q/2,2Q]$). Fixed-order perturbative calculations using conventional scale setting are subject to renormalization scale and scheme ambiguities~\cite{Celmaster:1979km,Abbott:1980hwa,Buras:1979yt,Grunberg:1980ja,Stevenson:1981vj,Brodsky:1982gc}. An additional theoretical ambiguity arises from the two typical scales (the Higgs boson mass $m_H$ and the $J/\psi$ meson mass $m_{J/\psi}$) involved in the $H \to J/\psi+\gamma$ process. The common assumption that the renormalization scale is based on the Higgs boson mass $m_H$, rather than on the $J/\psi$ meson mass $m_{J/\psi}$, lacks clear justification. Simply selecting the renormalization scale using conventional scale setting results in large logarithms of the type $\ln(m^2_H/m^2_{J/\psi})$ in the QCD corrections, which compromise the perturbative results. To restore the correct behavior of the perturbative coefficients, resummation techniques must be introduced and combined with the NRQCD and light-cone formalisms~\cite{Lepage:1980fj,Chernyak:1983ej,Jia:2008ep}. Thus, to obtain precise predictions for the $H \to J/\psi+\gamma$ process, it is essential to adopt a reliable method to eliminate the renormalization scale ambiguity.

The Principle of Maximum Conformality (PMC) provides an unambiguous and systematic approach to eliminating the renormalization scheme-and-scale ambiguities in perturbative QCD predictions~\cite{Brodsky:2011ta,Brodsky:2012rj,Brodsky:2011ig,Mojaza:2012mf,Brodsky:2013vpa,DiGiustino:2023jiq,Huang:2021hzr}. The PMC method is the underlying principle for the well-known Brodsky-Lepage-Mackenzie (BLM) method~\cite{Brodsky:1982gc} and is identical to the BLM method at next-to-leading order, providing a reliable extension of the BLM method to all orders. The BLM and PMC reduce to the Gell-Mann-Low method~\cite{Gell-Mann:1954yli} in the Abelian limit~\cite{Brodsky:1997jk,Wang:2020ckr,DiGiustino:2021nep}. The essential step in the PMC procedure is to identify all contributions originating from the $\beta$-terms in a pQCD series; one then shifts the renormalization scales of the QCD running coupling at each order to absorb the $\beta$-terms. The resulting scale-fixed perturbative series is thus identical to the corresponding "conformal" series, which has $\beta=0$; the PMC prediction possesses the essential feature of being renormalization-scheme-independent at every finite order and satisfies the principle of renormalization group invariance (RGI)~\cite{Brodsky:2012ms,Wu:2014iba,Wu:2018cmb,Wu:2019mky}. Due to the elimination of all $\beta$ terms, which include those related to factorial renormalon divergence, the convergence of the pQCD series is improved.

The PMC method has been successfully applied to several Higgs physics processes, leading to improvements in the precision of theoretical predictions. Results from the PMC application to the Higgs production process at the LHC~\cite{Wang:2016wgw} have shown improved agreement with ATLAS measurements compared to previous calculations obtained using the conventional scale-setting method. Detailed PMC analyses for the Higgs boson decay channels $H \to b\bar{b}$ and $H \to gg$ can be found in Refs.~\cite{Wang:2013bla, Zeng:2015gha, Zeng:2018jzf, Gao:2021wjn, Yan:2022foz, Yan:2024oyb}. The improved PMC predictions for the $H \to \gamma\gamma$ decay process are presented in Ref~\cite{Wang:2013akk, Yu:2018hgw}, demonstrating that the PMC-determined scale significantly differs from the typical process scale $\mu_r=m_H$. Moreover, these analyses have provided more reliable estimates of the unknown higher-order QCD corrections for these physical processes.

Achieving reliable and precise predictions for the $H \to J/\psi+\gamma$ process is essential for further exploring the Higgs particle. In this paper, we extend our previous PMC analysis of Higgs physics by performing calculations for the decay process $H \to J/\psi+\gamma$ using the PMC method. The remaining sections of this paper are organized as follows. In Sec.\ref{sec:2}, we present the calculation techniques for the Higgs decay $H \to J/\psi+\gamma$ up to NNLO QCD corrections using the PMC scale setting. In Sec.\ref{sec:3}, we present the numerical results and discussions for the Higgs decay $H \to J/\psi+\gamma$. Section \ref{sec:4} is reserved for a summary.

\section{PMC scale setting for the $H \to J/\psi+\gamma $ process}
\label{sec:2}

In the $H \to J/\psi+\gamma$ process, the decay width depends significantly on the charm quark Yukawa coupling~\cite{Bodwin:2013gca}. The $J/\psi$ production is facilitated by two distinct mechanisms. In the direct channel, the Higgs boson initially decays into a color-singlet $c\bar{c}$ pair. Subsequently, this pair decays into the $J/\psi$ while emitting a hard photon. This mechanism was first discussed in Ref.~\cite{Keung:1983ac}. The indirect channel occurs at least at the one-loop level. Here, the Higgs boson initially decays into a real photon and a virtual photon via either a $W$ boson or a heavy fermion, followed by the fragmentation of the virtual photon into $J/\psi$. Calculations for the $H \to J/\psi+\gamma$ process are presented in Ref.~\cite{Jia:2024ini}. Predictions for the indirect channel indicate that the higher-order QCD corrections have a negligible effect and are relatively insensitive to the renormalization scale $\mu_r$, as well as to the masses of the $c$ and $b$ quarks. In contrast, for the direct channel, the one-loop and two-loop corrections are negative and substantial; these impact the convergence of the pQCD series, potentially causing it to be ambiguous. Consequently, contributions to the direct channel exhibit a greater $\mu_r$-scale dependence than those for the indirect channel. Therefore, in this paper, we provide a detailed PMC analysis for the direct channel.

The total decay rate for the $H \to J/\psi+\gamma$ process can be organized as follows:
\begin{eqnarray}
\Gamma&=&\frac{(m_H^2 - m_{J/\psi}^2)^3}{8 \pi\,m_H^3} \biggl| F_{\text{dir}}(\tau) + F_{\text{ind}}(\tau) \biggl|^2,
\label{EQ:totdecay}
\end{eqnarray}
where $m_{H}$ is the Higgs-boson mass, $m_{J/\psi}$ is the $J/\psi$ meson mass, and $F_\mathrm{dir}(\tau)$ and $F_\mathrm{ind}(\tau)$ (with $\tau={4 m_{c}^{2}}/{m_{H}^{2}}$) represent the contributions for the direct and indirect $J/\psi$ production mechanisms, respectively.

The NNLO QCD corrections for the direct channel are given by:
\begin{eqnarray}
F_{\text{dir}}=A_{\rm{LO}}\bigg[1+c_1\,a_s(\mu_r)+c_2(\mu_r)\,a_s^2(\mu_r)+\mathcal O(a_s^3)\bigg].
\label{EQ:fdir}
\end{eqnarray}
Here, $a_s(\mu_r)=\alpha_s(\mu_r)/4\pi$, and the leading-order (LO) correction is
\begin{eqnarray}
A_{\rm{LO}} = \sqrt{\frac{G_F}{\sqrt{2}}} \cdot \frac{4 ee_c m_c}{m_H^2} \cdot \sqrt{\frac{N_c}{2\pi}} R_{J/\psi}(0) \cdot \frac{1}{1 - \tau},
\end{eqnarray}
where $G_{F}$ is the Fermi constant, $ee_{c}=\frac{2}{3}e$ denotes the fractional electric charge of the $c$-quark, $m_{c}$ is the $c$-quark mass, $N_c=3$ is the number of colors, and $R_{J/\psi}(0)$ is the radial Schrödinger wave function of $J/\psi$ at the origin.

The coefficients $c_1$ and $c_2(\mu_r)$ represent the next-to-leading order (NLO) and next-to-next-to-leading order (NNLO) perturbative QCD (pQCD) corrections, respectively. The NNLO coefficient $c_2(\mu_r)$ is composed of two components: one that depends on the number of active quark flavors, $n_f$, and one that is independent of it, i.e.,
\begin{eqnarray}
c_2(\mu_r)=c_{2,0}(\mu_r)+c_{2,1}(\mu_r)\,n_{f}.
\end{eqnarray}
The relation between the number of active quark flavors, denoted as $n_f$, and the $\beta_{0}$ term is expressed by the equation $\beta_{0}=11-\frac{2}{3}\,n_f$. The NNLO pQCD corrections for the direct channel in Eq. (\ref{EQ:fdir}) can be reformulated as
\begin{eqnarray}
F_{\text{dir}}&=&A_{\rm{LO}}\biggl[1+r_{1,0}\,a_s(\mu_r)+(r_{2,0}(\mu_r)  \nonumber \\
&& +r_{2,1}(\mu_r)\cdot\beta_0)\,a_s^2(\mu_r) + \mathcal O(a_s^3) \biggl],
\end{eqnarray}
where the coefficients $r_{1,0}$ and $r_{2,0}(\mu_r)$ represent the conformal coefficients, while the coefficient $r_{2,1}(\mu_r)$ signifies the non-conformal coefficient. These coefficients are
\begin{eqnarray}
r_{1,0}&=&c_1, \\
r_{2,0}(\mu_r)&=&c_{2,0}(\mu_r)+\frac{33}{2}\,c_{2,1}(\mu_r), \\
r_{2,1}(\mu_r)&=&-\frac{3}{2}\,c_{2,1}(\mu_r).
\end{eqnarray}

We adopt the PMC single-scale method~\cite{Shen:2017pdu} for the analysis. By integrating the non-conformal coefficients into the QCD coupling constant for the direct channel correction, we obtain
\begin{eqnarray}
F_{\text{dir}}=A_{\rm{LO}}\biggl[1+r_{1,0}\,a_s(Q_\star)+r_{2,0}(\mu_r)\,a_s^2(Q_\star) + \mathcal O(a_s^3) \biggr],
\label{EQ:PMCfdir}
\end{eqnarray}
where the PMC scale $Q_\star$ is
\begin{eqnarray}
Q_\star=\mu_r\,\exp\left[-\frac{r_{2,1}(\mu_r)}{2\,r_{\rm 1,0}}\right].
\label{PMCscaleQ}
\end{eqnarray}
At the NNLO level, only the $\beta_0$ term appears. This $\beta_0$ term, along with the scale-dependent logarithmic term, is resummed into the strong coupling $\alpha_s$ using the PMC method. The renormalization scale $\mu_r$ dependence in both the PMC scale $Q_\star$ and the conformal coefficients $r_{2,0}(\mu_r)$ is completely eliminated, resulting in a precise theoretical prediction that is independent of any $\mu_r$ variation.

\section{Numerical results and discussions}
\label{sec:3}

For numerical calculations, we utilize the following parameter values~\cite{Jia:2024ini}: $G_F=1.1664\times 10^{-5}\,\mathrm{GeV}^{-2}$, $m_H=125.20\,\mathrm{GeV}$, $m_{J/\psi}=3.0969\,\mathrm{GeV}$, $m_c=1.5\,\mathrm{GeV}$, and $R_{J/\psi}(0)=\sqrt{0.81\,\text{GeV}^{3}}$~\cite{Eichten:1995ch}. We also implement the two-loop $\alpha_s$ running coupling with $\alpha_s(M_Z)=0.1180$~\cite{ParticleDataGroup:2024cfk}.

It is important to note that, in the indirect channel, the decay of the Higgs boson is mediated by processes involving the $W$ boson or heavy fermions, which are operative at the one-loop level. We obtain
\begin{eqnarray}
F_{\rm{ind}}|_{\rm{LO}}=-43.577\times10^{-8}\,\, \rm{GeV}^{-1},
\end{eqnarray}
```latex
for the one-loop contribution~\cite{Bergstrom:1985hp}; two-loop corrections, however, are negligible~\cite{Jia:2024ini}. These can be described as:
```
\begin{eqnarray}
F_{\rm{ind}}|_{\rm{NLO}}=-43.990\times10^{-8}\,\, \rm{GeV}^{-1},
\end{eqnarray}
which contributes only $0.9\%$ relative to the one-loop result. Notably, predictions for the indirect channel are insensitive to changes in the renormalization scale $\mu_r$ and are not affected by the masses of the $c$ and $b$ quarks~\cite{Jia:2024ini}.

In contrast, the direct channel exhibits significant negative corrections at both the one-loop and two-loop levels, suggesting potential issues with pQCD convergence. The direct channel also shows a greater sensitivity to the scale $\mu_r$ compared to the indirect channel. To address these challenges, we conducted calculations for the direct channel using both the conventional and PMC scale settings. We then integrated the contributions from both the direct and indirect channels, as described by Eq.(\ref{EQ:totdecay}), to derive the total decay width for the $H \to J/\psi+\gamma$ process.

Specifically, the contributions from the direct channel at LO, NLO, and NNLO using the conventional scale setting are provided by:
\begin{eqnarray}
F_{\rm{dir}}|_{\rm{LO}}&=&A_{\rm{LO}}, \\
F_{\rm{dir}}|_{\rm{NLO}}&=&A_{\rm{LO}}\bigg[1+c_1\,a_s(\mu_r)\bigg], \\
F_{\rm{dir}}|_{\rm{NNLO}}&=&A_{\rm{LO}}\bigg[1+c_1\,a_s(\mu_r)+c_2\,a_s^2(\mu_r)\bigg].
\end{eqnarray}
The corresponding PMC results can be obtained by substituting $c_1$, $c_2$, and $\mu_r$ with $r_{1,0}$, $r_{2,0}$, and $Q_\star$, respectively. The decay widths for LO, NLO, and NNLO, denoted as $\Gamma_{\rm{LO}}$, $\Gamma_{\rm{NLO}}$, and $\Gamma_{\rm{NNLO}}$, are:
\begin{eqnarray}
\Gamma_{\rm{LO}}&=&\frac{(m_H^2 - m_{J/\psi}^2)^3}{8 \pi\,m_H^3} \biggl| F_{\rm{dir}}|_{\rm{LO}}+F_{\rm{ind}}|_{\rm{LO}}\biggl|^2, \\
\Gamma_{\rm{NLO}}&=&\frac{(m_H^2 - m_{J/\psi}^2)^3}{8 \pi\,m_H^3} \biggl| F_{\rm{dir}}|_{\rm{NLO}}+ F_{\rm{ind}}|_{\rm{NLO}} \biggl|^2, \\
\Gamma_{\rm{NNLO}}&=&\frac{(m_H^2 - m_{J/\psi}^2)^3}{8 \pi\,m_H^3} \biggl| F_{\rm{dir}}|_{\rm{NNLO}}+ F_{\rm{ind}}|_{\rm{NLO}} \biggl|^2.
\end{eqnarray}
The terms for NLO and NNLO QCD corrections, denoted as $\delta\Gamma_{\rm{NLO}}$ and $\delta\Gamma_{\rm{NNLO}}$, are expressed as follows:
\begin{eqnarray}
\delta\Gamma_{\rm{NLO}}&=&\Gamma_{\rm{NLO}}-\Gamma_{\rm{LO}}, \\
\delta\Gamma_{\rm{NNLO}}&=&\Gamma_{\rm{NNLO}}-\Gamma_{\rm{NLO}}.
\end{eqnarray}

The contributions for the direct channel using the conventional scale setting are:
\begin{eqnarray}
\Gamma_{\rm{dir}}&=&\frac{(m_H^2 - m_{J/\psi}^2)^3}{8 \pi\,m_H^3} \biggl| F_{\rm{dir}}\biggl|^2 \nonumber\\
&=&0.529-0.404-0.081=0.044, \\
&=&0.529-0.372-0.096=0.060, \\
&=&0.529-0.345-0.108=0.076,  \\
&=&0.529-0.322-0.116=0.091,
\end{eqnarray}
for $\mu_r = m_H/4$, $m_H/2$, $m_H$, and $2m_H$, respectively. The NLO and NNLO QCD corrections are both negative and substantial, which may obscure the convergence of the pQCD series. The quality of pQCD convergence exhibits significant dependence on the choice of $\mu_r$ scale. It remains indeterminate whether the poor convergence is an intrinsic characteristic of the pQCD series or results from an inappropriate choice of the scale $\mu_r$. The contributions for the direct channel using the PMC scale setting are
\begin{eqnarray}
\Gamma_{\rm{dir}}&=&\frac{(m_H^2 - m_{J/\psi}^2)^3}{8 \pi\,m_H^3} \biggl| F_{\rm{dir}}\biggl|^2 \nonumber\\
&=&0.529-0.523+0.008=0.014,
\end{eqnarray}
for any choice of scale $\mu_r$. The NNLO QCD correction is significantly suppressed, and the pQCD convergence is notably enhanced using the PMC scale setting compared to the conventional approach. For the $H \to J/\psi+\gamma$ process, although the decay width is primarily governed by the indirect channel contribution, the accurate PMC prediction for the direct channel enables a more precise theoretical estimation of the decay width.

\begin{table} [htb]
\begin{tabular}{|c|c|c|c|c|c|}
\hline
~~ ~~  & ~$\mu_r$~ & ~$\Gamma_{\rm LO}$~ & ~$\delta\Gamma_{\rm NLO}$~ & ~$\delta\Gamma_{\rm NNLO}$~ & ~$\Gamma_{\rm NNLO}$~ \\
\hline
~Conv.(eV)~  & ~$m_H/4$~ &~9.735~& ~2.726~ & ~1.028~ & ~13.490~  \\
~~       ~~  & ~$m_H/2$~ &~9.735~& ~2.433~ & ~1.075~ & ~13.243~  \\
~~       ~~  & ~$m_H$~   &~9.735~& ~2.203~ & ~1.081~ & ~13.020~  \\
~~       ~~  & ~$2m_H$~  &~9.735~& ~2.026~ & ~1.077~ & ~12.839~  \\
\hline
~PMC.(eV)~   & ~~        &~9.735~& ~4.778~ & ~-0.331~  & ~14.183~  \\
\hline
\end{tabular}
\caption{The NLO and NNLO QCD corrections, denoted as $\delta\Gamma_{\rm{NLO}}$ and $\delta\Gamma_{\rm{NNLO}}$, for the $H \to J/\psi+\gamma$ process are studied using both the conventional (Conv.) and the PMC scale setting methods. The factorization scale is set at $\mu_\Lambda=m_c=1.5$ GeV. For comparative purposes, the LO decay width $\Gamma_{\rm{LO}}$ for the $H \to J/\psi+\gamma$ process is also presented.  }
\end{table}

The pQCD corrections, $\delta\Gamma_{\rm{NLO}}$ and $\delta\Gamma_{\rm{NNLO}}$, for the $H \to J/\psi+\gamma$ process are presented in Table \ref{tab1}. A comparison of these corrections with the LO decay width is also provided in Table \ref{tab1}. It is observed that the LO decay width is dominant and unaffected by strong interactions. Perturbative QCD effects first appear at the NLO. In the conventional scale setting, the renormalization scale is chosen to be equal to the Higgs boson mass and is independent of the charm quark mass $m_c$ (e.g., $\mu_r=m_H/2=Q$). The conventional uncertainty is assessed by varying the scale parameter within a predetermined range (such as $\mu_r \in [Q/2, 2Q]$).

The relative impact of the perturbative corrections on the LO decay width is
\begin{equation} \frac{\delta\Gamma_{\rm{NLO}}}{\Gamma_{\rm{LO}}}:\frac{\delta\Gamma_{\rm{NNLO}}}{\Gamma_{\rm{LO}}} \sim22.6\%:11.1\%
\end{equation}
for $\mu_r=m_H$. These values change to approximately 25.0\%:11.0\% at the scale $\mu_r=m_H/2$, and to about 28.0\%:10.6\% at $\mu_r=m_H/4$. We observe improved convergence of the pQCD series at lower scale values and slower convergence at higher scales. Therefore, the convergence quality of the pQCD series is significantly influenced by the choice of the renormalization scale. Using the conventional scale setting, the scale uncertainties are evaluated by varying the scale within the range $\mu_r\in[Q/2,2Q]$. When the scale $Q=m_H/2$ is used as the central value, the scale uncertainties are $\left(^{+12.0\%}_{-9.5\%}\right)$ and $\left(^{+0.6\%}_{-4.4\%}\right)$ for $\delta\Gamma_{\rm NLO}$ and $\delta\Gamma_{\rm NNLO}$, respectively.

The NLO QCD correction decreases, whereas the NNLO QCD correction increases with the scale $\mu_r$. The scale dependence cancels between the NLO and NNLO terms. Consequently, the scale uncertainty is reduced when NNLO QCD corrections are included.

The NLO total decay widths are $\Gamma^{\rm Conv}_{\rm NLO}=12.462, 12.169, 11.938, 11.761$ eV for the values $\mu_r=m_H/4$, $m_H/2$, $m_H$, and $2m_H$, respectively. Thus, using the conventional approach, the uncertainty is $\left(^{+2.4\%}_{-1.9\%}\right)$ when the scale is set to $Q=m_H/2$. Results for the total decay width at NNLO are $\Gamma^{\rm Conv}_{\rm NNLO}=13.490, 13.243, 13.020, 12.839$ eV for $\mu_r=m_H/4$, $m_H/2$, $m_H$, and $2m_H$, respectively. The corresponding scale uncertainty is $\left(^{+1.9\%}_{-1.7\%}\right)$ for $Q=m_H/2$.

Using the PMC scale setting, the non-conformal $\beta_0$-term and the scale-dependent logarithmic term are resummed into the strong coupling $\alpha_s$. By employing Eq.(\ref{PMCscaleQ}), we obtain the value:
\begin{eqnarray}
Q_\star=3.29 \,\,\rm{GeV},
\end{eqnarray}
for the PMC scale, which is independent of the renormalization scale $\mu_r$.

The PMC scale is significantly lower than the Higgs boson mass $m_H$, indicating the limited virtuality of the QCD dynamics in the $H \to J/\psi+\gamma$ process. The effective momentum flow for this process should be $\mu_r \ll m_H$, which is more aligned with the charm quark mass ($2m_c$) than with the Higgs boson mass $m_H$. Importantly, if the conventional scale setting is applied and a smaller scale $\mu_r \ll m_H$ is selected, the pQCD convergence for the $H \to J/\psi+\gamma$ process is significantly enhanced. This leads to predictions that closely align with those derived using the PMC.

Apart from the scale of $\alpha_s$, the NNLO coefficients for the direct channel also differ substantially between the conventional scale setting and the PMC approach. When using the conventional scale setting, the NNLO coefficient is:
\begin{eqnarray}
c_2(\mu_r)=-2092.236\pm529.425,
\end{eqnarray}
where the error is determined by the variation of the scale $\mu_r\in[Q/2, 2Q]$, with $Q=m_H/2$. The NNLO coefficient $c_2(\mu_r)$, using conventional scale setting, exhibits a pronounced dependence on the renormalization scale $\mu_r$.

In contrast, the scale-independent NNLO conformal coefficient $r_{2,0}(\mu_r)$, implemented via the PMC, is expressed as:
\begin{eqnarray}
r_{2,0}(\mu_r)=156.244
\end{eqnarray}
which is significantly smaller in magnitude compared to the NNLO conventional coefficient $c_2(\mu_r)$. It is observed that the NNLO coefficient is negative under the conventional scale setting, but becomes positive when using the PMC.

For any chosen renormalization scale $\mu_r$, the PMC NLO QCD correction term is consistently $4.778$ eV, and the PMC NNLO QCD correction term is consistently $-0.331$ eV. The relative impact of the PMC correction on the width at LO order is:
\begin{eqnarray}
\frac{\delta\Gamma_{\rm{NLO}}}{\Gamma_{\rm{LO}}}:\frac{\delta\Gamma_{\rm{NNLO}}}{\Gamma_{\rm{LO}}}\sim 49.1\%:-3.4\%.
\end{eqnarray}
Comparing the results, we observe that the PMC NLO QCD correction term is significantly increased, while the PMC NNLO QCD correction term is diminished compared to conventional results. This phenomenon primarily arises from the cancellation of non-conformal $\beta$ terms in the PMC pQCD series. Consequently, the final total decay width becomes independent of the choice of the renormalization scale $\mu_r$, resulting in the output
\begin{eqnarray}
\Gamma^{\rm PMC}_{\rm NNLO}=14.183 \,\,\rm{eV}.
\end{eqnarray}

\begin{figure}
\centering
\includegraphics[width=0.4\textwidth]{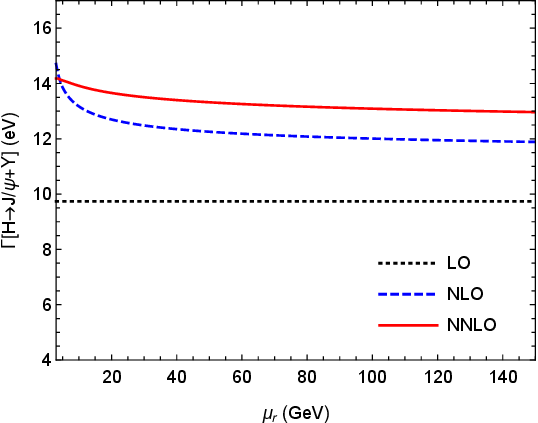}
\caption{The renormalization scale ($\mu_r$) dependence of the decay widths for the $H \to J/\psi+\gamma$ process is analyzed at LO, NLO, and NNLO using the conventional scale setting. The factorization scale is fixed at $\mu_\Lambda = m_c = 1.5$ GeV.}
\label{Fig:Convdw}
\end{figure}

We present the renormalization scale $\mu_r$ dependence of the decay width for the $H \to J/\psi+\gamma$ process at LO, NLO, and NNLO using conventional scale setting in Fig. (\ref{Fig:Convdw}). In the region of a large scale $\mu_{r}$, the decay width exhibits a weak dependence on $\mu_{r}$, whereas in the region of a small scale $\mu_{r}$, it shows a strong dependence on $\mu_{r}$. Additionally, the inclusion of NNLO QCD corrections further reduces the scale dependence. This observation aligns with the typical expectation of scale dependence cancellation as the level of accuracy in the fixed-order calculation improves. Figure (\ref{Fig:Convdw}) clearly demonstrates that varying the scale within the range $\mu_r \in [Q/2, 2Q]$ results in the NNLO prediction not overlapping with the NLO results. This inconsistency suggests a discrepancy in the correct and reliable assessment of the uncalculated higher-order contributions in QCD. In pQCD, the renormalization scale appears in non-conformal contributions (associated with the $\beta$-terms), but not in conformal terms. Indeed, the correct scale and its appropriate range of variation are unknown, making it challenging to obtain reliable quantitative predictions for theoretical uncertainties using the conventional scale setting.

\begin{figure}
\centering
\includegraphics[width=0.4\textwidth]{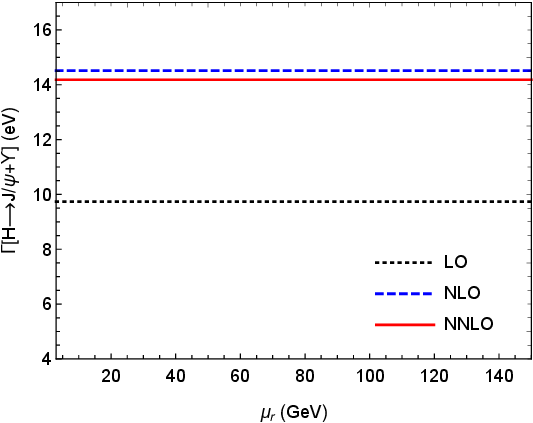}
\caption{The dependence of the renormalization scale ($\mu_r$) on the decay widths for the $H \to J/\psi + \gamma$ process is examined at LO, NLO, and NNLO using the PMC scale setting. The factorization scale is set at $\mu_\Lambda = m_c = 1.5$ GeV.}
\label{Fig:PMCdw}
\end{figure}

In Fig.~(\ref{Fig:PMCdw}), we present the dependence of the renormalization scale, $\mu_r$, on the decay widths for the $H \to J/\psi+\gamma$ process at LO, NLO, and NNLO using the PMC scale setting. Figure~(\ref{Fig:PMCdw}) indicates that the decay width for the $H \to J/\psi+\gamma$ process at NLO or NNLO remains nearly unchanged with variations in the renormalization scale $\mu_r$. The scale dependence associated with each perturbative order and with the total width is eliminated simultaneously. Consequently, the decay width at NLO or NNLO is no longer dependent on the renormalization scale. In comparison to the conventional scale-setting approach, the predictions obtained using the PMC method exhibit improved convergence of the perturbative series.

Once the renormalization scale ambiguity is eliminated, other sources of uncertainty remain, such as the factorization scale $\mu_\Lambda$ and unknown higher-order QCD contributions. The factorization scale ambiguity also affects conformal theories and could be eliminated by matching perturbative predictions with nonperturbative bound-state dynamics~\cite{Brodsky:2014yha}. With the introduction of the PMC, the decay width $\Gamma^{\rm{PMC}}_{\rm NNLO}$ increases with the factorization scale. The results for the decay width are $\Gamma^{\rm{PMC}}_{\rm NNLO}=13.836$, $14.183$, and $14.432$ eV for $\mu_\Lambda=1$, $1.5$, and $2$ GeV, respectively. With the PMC removing the renormalization scale ambiguity, a different method for estimating contributions from unknown higher-order QCD corrections must be introduced. By adopting the PMC scale setting, the scale is unambiguously determined by non-conformal contributions, and the conventional uncertainty estimate by varying the scale in $\alpha_s$ is not applicable within the PMC framework, as it would violate renormalization group invariance~\cite{Wu:2014iba}.

The PMC scale $Q_\star$ itself is a perturbative expansion series in $\alpha_s$ and can be determined up to leading-logarithm (LL) and next-to-leading-logarithm (NLL) accuracy using non-conformal contributions at NNLO and next-to-next-to-next-to-leading order (NNNLO), respectively. Generally, the perturbative series of the PMC scales $Q_\star$ demonstrate rapid pQCD convergence, and the inclusion of higher-order QCD contributions only slightly alters the PMC scales $Q_\star$ at LL. Due to unknown higher-order QCD contributions, there is residual scale dependence for the PMC scale. However, this residual scale dependence is highly suppressed and distinct from conventional scale ambiguities. We apply the Padé approximant approach (PAA)~\cite{Basdevant:1972fe, Samuel:1992qg, Samuel:1995jc} to the PMC results to estimate unknown higher-order QCD contributions for the $H \to J/\psi+\gamma$ process.

The Padé approximant approach provides an estimate of the magnitude of the unknown $(n+1)_{\rm th}$-order term, starting from the calculated $n_{\rm th}$-order perturbative series. The $[N/M]$-type approximant, denoted as $\rho_n^{[N/M]}$, for the observable $\rho_n=\sum_{i=0}^{n(\geq 1)}C_i x^i$ calculated up to the $n_{\rm th}$-order, is defined by:
\begin{eqnarray}
\rho^{[N/M]}_n &=& \frac{b_0+b_1 x + \cdots + b_N x^N} {1 + c_1 x + \cdots + c_M x^M} \label{eq:rhonm1} \\
               &=& \sum_{i=0}^{n} C_i x^i + C_{n+1} \; x^{n+1} +\cdots,
               \label{eq:rhonm}
\end{eqnarray}
where the parameter \( M \geq 1 \) and \( N + M = n \). The known coefficients \( C_{i (\leq n)} \) determine the values of the parameters \( b_{i \in [0, N]} \) and \( c_{j \in [1, M]} \) up to the \( n_{\text{th}} \)-order. The coefficient \( C_{n+1} \) in Eq. (\ref{eq:rhonm}), i.e., the coefficient of the next uncalculated order, is obtained by performing a Taylor expansion of the rational function in Eq. (\ref{eq:rhonm1}) up to the \( (n+1)_{\text{th}} \) order of accuracy~\cite{Du:2018dma}.

Applying the PAA method to Eq. (9), the estimated NNLO coefficient is \( r_{2,0}^{\rm PAA} = r_{1,0}^2 \) for \( [N/M] = [0/1] \); the estimated NNNLO coefficient is \( r_{3,0}^{\rm PAA} = r_{2,0}^2/r_{1,0} \) for \( [N/M] = [1/1] \). Thus, the uncertainties from unknown higher-order terms can be estimated as \( \pm r_{2,0}^{\rm PAA}\,a_s^2(Q_\star) \) for the NLO PMC result and \( \pm r_{3,0}^{\rm PAA}\,a_s^3(Q_\star) \) for the NNLO PMC result. The corresponding errors for the decay widths are \( \Delta\Gamma(H\to J/\psi+\gamma)|_{\rm{NLO}}=^{+4.814}_{-4.126} \) eV at NLO and \( \Delta\Gamma(H\to J/\psi+\gamma)|_{\rm{NNLO}}=\pm0.022 \) eV at NNLO. Since the PMC NLO QCD correction term is enhanced while the PMC NNLO QCD correction is significantly suppressed, the error of the estimated NNLO result is large, and the error arising from the unknown NNNLO result is significantly reduced. The exact NNLO PMC result is well within the error bar of the estimated NNLO result based on the PMC+PAA calculation. Finally, using the total Higgs width \( \Gamma_H = 4.07^{+4.0\%}_{-3.9\%} \, \text{MeV} \) provided in Refs.~\cite{CMS:2019ekd,Passarino:2016pzb} and the PMC results for the scale-independent decay width, we obtain a precise prediction for the branching ratio of the \( H \to J/\psi+\gamma \) decay process:
\begin{eqnarray}
\mathcal{B}(H \to J/\psi + \gamma) =3.485_{-0.161}^{+0.152} \times10^{-6}.
\end{eqnarray}

\section{Summary}
\label{sec:4}

In this paper, we conduct a comprehensive PMC analysis of the $H \to J/\psi+\gamma$ decay process. Under conventional scale setting, renormalization scale uncertainty emerges as a significant source of error for theoretical pQCD predictions, and the pQCD series shows slow convergence, particularly for the direct channel. The renormalization scale also influences the convergence of the pQCD series. Therefore, it is challenging to determine whether the slow convergence of the pQCD series is an intrinsic characteristic or a result of the specific renormalization scale choice. Furthermore, estimating unknown higher-order QCD contributions by varying the scale $\mu_r\in[Q/2,2Q]$ provides unreliable results. Specifically, the NNLO prediction does not coincide with the NLO results for the $H \to J/\psi+\gamma$ decay process.

By applying the PMC scale setting, the renormalization scale is unambiguously defined by incorporating all non-conformal terms, leaving only the conformal contributions in the perturbation series. Consequently, the renormalization scale uncertainty for the $H \to J/\psi + \gamma$ process is resolved. The PMC results in the following scale:
\begin{eqnarray}
Q_\star=3.29\,\,\rm{GeV},
\end{eqnarray}
which is an order of magnitude smaller than the conventional choice of $\mu_r=m_H/2$, reflecting the low virtuality of the underlying QCD dynamics for the $H \to J/\psi+\gamma$ process. The perturbative series obtained using the PMC exhibits improved convergence. The predicted decay width of the $H \to J/\psi + \gamma$ process is $\Gamma(H\to J/\psi+\gamma)=14.183^{+0.249}_{-0.347}\pm0.022$ eV, where the first error arises from the factorization scale $\mu_\Lambda \in [1, 2]$ GeV, and the second error stems from the estimation of the unknown higher-order terms using the PAA method. Moreover, we obtain a precise prediction for the branching ratio for the $H \to J/\psi + \gamma$ process, $\mathcal{B}(H \to J/\psi + \gamma) =3.485_{-0.161}^{+0.152} \times10^{-6}$. PMC predictions for the $H \to J/\psi + \gamma$ process can facilitate high-precision measurements at the High-Luminosity LHC (HL-LHC) and at future colliders such as the Future Circular Collider (FCC) and the Circular Electron Positron Collider (CEPC).

\hspace{1cm}

{\bf Acknowledgements}: This research was supported in part by the Natural Science Foundation of China under Grant Nos. 12265011, 12175025, and 12347101; by the Project of Guizhou Provincial Department of Science and Technology under Grant Nos. YQK[2023]016, ZK[2023]141, and CXTD[2025]030; by the Hunan Provincial Natural Science Foundation with Grant No. 2024JJ3004; by the YueLuShan Center for Industrial Innovation (2024YCII0118); and by the Natural Science Foundation of Sichuan Province under Grant No. 2024NSFSC1367.


\begin{thebibliography}{99}

\bibitem{higgs1}
G. Aad {\it et al.} [ATLAS Collaboration],
Phys. Lett. B {\bf 716}, 1 (2012).

\bibitem{higgs2}
S. Chatrchyan {\it et al.} [CMS Collaboration],
Phys. Lett. B {\bf 716}, 30 (2012).

\bibitem{higgsmass1}
[ATLAS Collaboration],
ATLAS-CONF-2013-014 (2013).

\bibitem{sign2}
[CMS Collaboration],
CMS-PAS-HIG-13-005 (2013).

\bibitem{sign1}
[ATLAS Collaboration],
ATLAS-CONF-2013-034 (2013).

\bibitem{cms131}
V.~Khachatryan {\it et al.} [CMS Collaboration],
Eur. Phys. J. C \textbf{74}, 3076 (2014).

\bibitem{atlas131}
[ATLAS Collaboration],
ATLAS-CONF-2013-012 (2013).

\bibitem{atlascms131}
[ATLAS Collaboration],
ATLAS-CONF-2013-013 (2013).

\bibitem{atlascms13002}
S.~Chatrchyan \textit{et al.} [CMS],
Phys. Rev. D \textbf{89}, 092007 (2014).

\bibitem{Kagan:2014ila}
A.~L.~Kagan, G.~Perez, F.~Petriello, Y.~Soreq, S.~Stoynev and J.~Zupan,
Phys. Rev. Lett. \textbf{114}, 101802 (2015).

\bibitem{CMS:2022psv}
A.~Tumasyan \textit{et al.} [CMS],
Phys. Rev. Lett. \textbf{131}, 061801 (2023).

\bibitem{ATLAS:2022rej}
G.~Aad \textit{et al.} [ATLAS],
Eur. Phys. J. C \textbf{83}, 781 (2023).

\bibitem{Shifman:1980dk}
M.~A.~Shifman and M.~I.~Vysotsky,
Nucl. Phys. B \textbf{186}, 475-518 (1981).

\bibitem{Vysotsky:1980cz}
M.~I.~Vysotsky,
Phys. Lett. B \textbf{97}, 159-162 (1980).

\bibitem{Keung:1983ac}
W.~Y.~Keung,
Phys. Rev. D \textbf{27}, 2762 (1983).

\bibitem{Bodwin:1994jh}
G.~T.~Bodwin, E.~Braaten and G.~P.~Lepage,
Phys. Rev. D \textbf{51}, 1125-1171 (1995)
[erratum: Phys. Rev. D \textbf{55}, 5853 (1997)]

\bibitem{Bodwin:2013gca}
G.~T.~Bodwin, F.~Petriello, S.~Stoynev and M.~Velasco,
Phys. Rev. D \textbf{88}, 053003 (2013).

\bibitem{Bodwin:2014bpa}
G.~T.~Bodwin, H.~S.~Chung, J.~H.~Ee, J.~Lee and F.~Petriello,
Phys. Rev. D \textbf{90}, 113010 (2014).

\bibitem{Konig:2015qat}
M.~K{\"o}nig and M.~Neubert,
JHEP \textbf{08}, 012 (2015).

\bibitem{Bodwin:2016edd}
G.~T.~Bodwin, H.~S.~Chung, J.~H.~Ee and J.~Lee,
Phys. Rev. D \textbf{95}, 054018 (2017).

\bibitem{Zhou:2016sot}
C.~Zhou, M.~Song, G.~Li, Y.~J.~Zhou and J.~Y.~Guo,
Chin. Phys. C \textbf{40}, 123105 (2016).

\bibitem{Bodwin:2017wdu}
G.~T.~Bodwin, H.~S.~Chung, J.~H.~Ee and J.~Lee,
Phys. Rev. D \textbf{96}, 116014 (2017).

\bibitem{Brambilla:2019fmu}
N.~Brambilla, H.~S.~Chung, W.~K.~Lai, V.~Shtabovenko and A.~Vairo,
Phys. Rev. D \textbf{100}, 054038 (2019).

\bibitem{Jia:2024ini}
Y.~Jia, Z.~Mo and J.~Y.~Zhang,
[arXiv:2408.17448 [hep-ph]].

\bibitem{Celmaster:1979km}
W.~Celmaster and R.~J.~Gonsalves,
Phys. Rev. D \textbf{20}, 1420 (1979).

\bibitem{Abbott:1980hwa}
L.~F.~Abbott,
Phys. Rev. Lett. \textbf{44}, 1569 (1980).

\bibitem{Buras:1979yt}
A.~J.~Buras,
Rev. Mod. Phys. \textbf{52}, 199 (1980).

\bibitem{Grunberg:1980ja}
G.~Grunberg,
Phys. Lett. B \textbf{95}, 70 (1980).

\bibitem{Stevenson:1981vj}
P.~M.~Stevenson,
Phys. Rev. D \textbf{23}, 2916 (1981).

\bibitem{Brodsky:1982gc}
S.~J.~Brodsky, G.~P.~Lepage and P.~B.~Mackenzie,
Phys. Rev. D \textbf{28}, 228 (1983).

\bibitem{Lepage:1980fj}
G.~P.~Lepage and S.~J.~Brodsky,
Phys. Rev. D \textbf{22}, 2157 (1980).

\bibitem{Chernyak:1983ej}
V.~L.~Chernyak and A.~R.~Zhitnitsky,
Phys. Rept. \textbf{112}, 173 (1984).

\bibitem{Jia:2008ep}
Y.~Jia and D.~Yang,
Nucl. Phys. B \textbf{814}, 217-230 (2009).

\bibitem{Brodsky:2011ta}
S.~J.~Brodsky and X.~G.~Wu,
Phys. Rev. D \textbf{85}, 034038 (2012)
[erratum: Phys. Rev. D \textbf{86}, 079903 (2012)]

\bibitem{Brodsky:2012rj}
S.~J.~Brodsky and X.~G.~Wu,
Phys. Rev. Lett. \textbf{109}, 042002 (2012).

\bibitem{Brodsky:2011ig}
S.~J.~Brodsky and L.~Di Giustino,
Phys. Rev. D \textbf{86}, 085026 (2012).

\bibitem{Mojaza:2012mf}
M.~Mojaza, S.~J.~Brodsky and X.~G.~Wu,
Phys. Rev. Lett. \textbf{110}, 192001 (2013).

\bibitem{Brodsky:2013vpa}
S.~J.~Brodsky, M.~Mojaza and X.~G.~Wu,
Phys. Rev. D \textbf{89}, 014027 (2014).

\bibitem{DiGiustino:2023jiq}
L.~Di Giustino, S.~J.~Brodsky, P.~G.~Ratcliffe, X.~G.~Wu and S.~Q.~Wang,
Prog. Part. Nucl. Phys. \textbf{135}, 104092 (2024).

\bibitem{Huang:2021hzr}
X.~D.~Huang, J.~Yan, H.~H.~Ma, L.~Di Giustino, J.~M.~Shen, X.~G.~Wu and S.~J.~Brodsky,
Nucl. Phys. B \textbf{989}, 116150 (2023).

\bibitem{Gell-Mann:1954yli}
M.~Gell-Mann and F.~E.~Low,
Phys. Rev. \textbf{95}, 1300-1312 (1954).

\bibitem{Brodsky:1997jk}
S.~J.~Brodsky and P.~Huet,
Phys. Lett. B \textbf{417}, 145-153 (1998).

\bibitem{Wang:2020ckr}
S.~Q.~Wang, S.~J.~Brodsky, X.~G.~Wu, L.~Di Giustino and J.~M.~Shen,
Phys. Rev. D \textbf{102}, 014005 (2020).

\bibitem{DiGiustino:2021nep}
L.~Di Giustino, F.~Sannino, S.~Q.~Wang and X.~G.~Wu,
Phys. Lett. B \textbf{823}, 136728 (2021).

\bibitem{Brodsky:2012ms}
S.~J.~Brodsky and X.~G.~Wu,
Phys. Rev. D \textbf{86}, 054018 (2012).

\bibitem{Wu:2014iba}
X.~G.~Wu, Y.~Ma, S.~Q.~Wang, H.~B.~Fu, H.~H.~Ma, S.~J.~Brodsky and M.~Mojaza,
Rept. Prog. Phys. \textbf{78}, 126201 (2015).

\bibitem{Wu:2018cmb}
X.~G.~Wu, J.~M.~Shen, B.~L.~Du and S.~J.~Brodsky,
Phys. Rev. D \textbf{97}, 094030 (2018).

\bibitem{Wu:2019mky}
X.~G.~Wu, J.~M.~Shen, B.~L.~Du, X.~D.~Huang, S.~Q.~Wang and S.~J.~Brodsky,
Prog. Part. Nucl. Phys. \textbf{108}, 103706 (2019).

\bibitem{Wang:2016wgw}
S.~Q.~Wang, X.~G.~Wu, S.~J.~Brodsky and M.~Mojaza,
Phys. Rev. D \textbf{94}, 053003 (2016).

\bibitem{Wang:2013bla}
S.~Q.~Wang, X.~G.~Wu, X.~C.~Zheng, J.~M.~Shen and Q.~L.~Zhang,
Eur. Phys. J. C \textbf{74}, 2825 (2014).

\bibitem{Zeng:2015gha}
D.~M.~Zeng, S.~Q.~Wang, X.~G.~Wu and J.~M.~Shen,
J. Phys. G \textbf{43}, 075001 (2016).

\bibitem{Zeng:2018jzf}
J.~Zeng, X.~G.~Wu, S.~Bu, J.~M.~Shen and S.~Q.~Wang,
J. Phys. G \textbf{45}, 085004 (2018).

\bibitem{Gao:2021wjn}
C.~T.~Gao, X.~G.~Wu, X.~D.~Huang and J.~Zeng,
Chin. Phys. C \textbf{46}, 123109 (2022).

\bibitem{Yan:2022foz}
J.~Yan, Z.~F.~Wu, J.~M.~Shen and X.~G.~Wu,
J. Phys. G \textbf{50}, 045001 (2023).

\bibitem{Yan:2024oyb}
J.~Yan, X.~G.~Wu, J.~M.~Shen, X.~D.~Huang and Z.~F.~Wu,
JHEP \textbf{04}, 184 (2025).

\bibitem{Wang:2013akk}
S.~Q.~Wang, X.~G.~Wu, X.~C.~Zheng, G.~Chen and J.~M.~Shen,
J. Phys. G \textbf{41}, 075010 (2014).

\bibitem{Yu:2018hgw}
Q.~Yu, X.~G.~Wu, S.~Q.~Wang, X.~D.~Huang, J.~M.~Shen and J.~Zeng,
Chin. Phys. C \textbf{43}, 093102 (2019).

\bibitem{Shen:2017pdu}
J.~M.~Shen, X.~G.~Wu, B.~L.~Du and S.~J.~Brodsky,
Phys. Rev. D \textbf{95}, 094006 (2017).

\bibitem{Eichten:1995ch}
E.~J.~Eichten and C.~Quigg,
Phys. Rev. D \textbf{52}, 1726-1728 (1995).

\bibitem{ParticleDataGroup:2024cfk}
S.~Navas \textit{et al.} [Particle Data Group],
Phys. Rev. D \textbf{110}, 030001 (2024).

\bibitem{Bergstrom:1985hp}
L.~Bergstrom and G.~Hulth,
Nucl. Phys. B \textbf{259}, 137-155 (1985)
[erratum: Nucl. Phys. B \textbf{276}, 744-744 (1986)].

\bibitem{Brodsky:2014yha}
S.~J.~Brodsky, G.~F.~de Teramond, H.~G.~Dosch and J.~Erlich,
Phys. Rept. \textbf{584}, 1-105 (2015).

\bibitem{Basdevant:1972fe}
J.~L.~Basdevant,
Fortsch. Phys. \textbf{20}, 283-331 (1972).

\bibitem{Samuel:1992qg}
M.~A.~Samuel, G.~Li and E.~Steinfelds,
Phys. Lett. B \textbf{323}, 188 (1994).

\bibitem{Samuel:1995jc}
M.~A.~Samuel, J.~R.~Ellis and M.~Karliner,
Phys. Rev. Lett. \textbf{74}, 4380-4383 (1995).

\bibitem{Du:2018dma}
B.~L.~Du, X.~G.~Wu, J.~M.~Shen and S.~J.~Brodsky,
Eur. Phys. J. C \textbf{79}, 182 (2019).

\bibitem{CMS:2019ekd}
A.~M.~Sirunyan \textit{et al.} [CMS],
Phys. Rev. D \textbf{99}, 112003 (2019).

\bibitem{Passarino:2016pzb}
G.~Passarino and M.~Trott,
[arXiv:1610.08356 [hep-ph]].



\end{thebibliography}
\end{document}